\documentclass[usenatbib]{mn2e}

\newcommand{\Msol}{\ensuremath{\mathrm{M_{\odot}\ }}}

\def\R200{\ensuremath{R_{\mathrm{200}\ }}}

\newcommand{\Zsol}{\ensuremath{\mathrm{Z_{\odot}}}}

\newcommand{\Chandra}{\emph{Chandra}\ }

\newcommand{\CIAO}{\emph{CIAO}\ }

\newcommand{\chisq}{\ensuremath{\chi^2}\ }
\newcommand{\rchisq}{\ensuremath{\chi^2_\nu}\ }

\newcommand{\nm}{\mbox{\ensuremath{\mathrm{~\nm}\ }}}

\newcommand{\s}{\mbox{\ensuremath{\mathrm{~s}\ }}}

\usepackage{subfigure,times,epsf, graphicx} 
\input{epsf}
\title[The shock in B2 0838+32A]{Shock heating in the group atmosphere
  of the radio galaxy B2 0838+32A}
\author[N.N. Jetha et al]{Nazirah
N. Jetha$^{1}$\thanks{E-mail:nazirah.jetha@cea.fr}, Martin
J. Hardcastle$^2$, Trevor J. Ponman$^3$ and Irini Sakelliou$^4$
\\$^{1}$Laboratoire AIM, CEA/DSM - CNRS - Universit\'{e} Paris
Diderot, DAPNIA/Service d'Astrophysique, B\^{a}t. 709, CEA-Saclay,
F-91191 Gif-sur-Yvette C\'{e}dex, France\\$^{2}$School of Physics,
Astronomy and Mathematics, University of Hertfordshire, College Lane,
Hatfield, Hertfordshire AL10 9AB\\$^3$School of Physics and Astronomy,
University of Birmingham, Edgbaston, Birmingham B15
2TT\\$^{4}$Max-Planck-Institute f\"{u}r Astronomie, K\"{o}nigstuhl 17,
D-69117 Heidelberg, Germany}

\begin{document} 
\date{} 
\maketitle{}
\label{firstpage} 
\begin{abstract}
We present {\it Chandra} and radio observations, and analysis of Sloan
Digital Sky Survey data, of the radio galaxy B2 0838+32A (4C\,32.26) and
its environment. The radio galaxy is at the centre of a nearby group
that has often been identified with the cluster Abell 695, but we
argue that the original Abell cluster is likely to be an unrelated and
considerably more distant system. The radio source is a restarting
radio galaxy and, using our {\it Chandra} data, we argue that the
currently active lobes are expanding supersonically, driving a shock
with Mach number $2.4^{+1.0}_{-0.5}$ into the inter-stellar medium.
This would be only the third strong shock round a young radio source
to be discovered, after Centaurus A and NGC 3801. However, in contrast
to both these systems, the host galaxy of B2 0838+32A shows no evidence
for a recent merger, while the AGN spectrum shows no evidence for the
dusty torus that would imply a large reservoir of cold gas close to
the central black hole. On the contrary, the AGN spectrum is of a type
that has been associated with the presence of a radiatively
inefficient accretion flow that could be controlled by AGN heating and
subsequent cooling of the hot, X-ray emitting gas. If correct, this
means that B2 0838+32A is the first source in which we can directly see
entropy-increasing processes (shocks) driven by accretion from the hot
phase of the interstellar medium.
\end{abstract}
\begin{keywords}\end{keywords}

\section{Introduction}
\label{introduction}
Recent observations of galaxy groups and clusters have highlighted two
major, and as yet unresolved, problems in our understanding of the formation
and evolution of these objects.  The first is that of similarity
breaking; whilst groups and clusters do follow well studied scaling
relations, they differ significantly from those derived from models of
self-similar gravitational collapse
(e.g. \citealt{2003MNRAS.340..989S}, \citealt{2005AA...433..431P}).
The second problem is that many of these systems exhibit cores of cool
gas with cooling times well below the Hubble time.  In theory, these
cool cores should have cooled to such temperatures that they are no
longer visible in the X-ray; however, no evidence of such cool gas is
found (e.g. \citealt{2001A&A...365L.104P},
\citealt{2002A&A...391..903S}).  The heating mechanism is still the
subject of vigorous debate, but there is increasing evidence that
active galactic nucleus (AGN) outbursts could heat the intergalactic
medium (IGM).  Depending on the scale of the outburst, not only could
they prevent catastrophic cooling in the core, but over repeated
cycles of activity outbursts may also inject sufficient extra energy
at larger radii to drive groups and clusters away from the predicted
self-similar scaling relations.

The heating on small scales can be achieved by shock heating of the
IGM by the rapidly expanding small-scale lobes of a radio-loud AGN
\citep{2003ApJ...592..129K}, whilst the larger scale heating may be
the result of the evolution of buoyant bubbles
(\citealt{2002MNRAS.332..271R}, \citealt{2006MNRAS.373..739N}, for
simulations; \citealt{2004ApJ...607..800B},
\citealt{2005ApJ...625..748C}, for observations). In the simulations,
overpressured bubbles of radio plasma that have evolved from old, dead
radio galaxies and which are light compared to the IGM, rise due to
buoyancy. As they rise, they expand and heat the IGM as the bubbles
come into pressure equilibrium
\citep[e.g.][]{2001ASPC..240..363B}. Alternatively, large scale
heating may also arise from the interactions between the radio jets
and the IGM \citep{2006MNRAS.373L..65H}, or from the dissipation of
large-scale shocks such as those seen in Perseus~A
\citep{2006MNRAS.366..417F} and Hydra~A \citep{2005ApJ...628..629N}.
The models therefore predict that we should expect to see several
different phases of AGN heating, and, in the rare sources in which
there is clear evidence for multiple AGN outbursts, we might expect to
see more than one such process operating at once.

However, the cumulative effect of AGN heating is not well understood
observationally, since there are very few systems that show evidence
of multiple AGN outbursts.  This makes examining the energy output of
a radio source at different stages of its evolution, whilst keeping
the external environment more or less consistent, rather
difficult. Finding evidence of multiple outbursts in the same system,
from the same host galaxy, allows us firstly to investigate the energy
output of the radio source at different stages of its lifetime and
secondly to determine how this energy can be coupled to the IGM at
different points in the radio galaxy's evolution.  Furthermore,
observing systems with multiple visible outbursts may provide
information as to how successive outbursts affect the IGM and how a
radio galaxy may, at different points in its evolution, perform
different functions; i.e. in the early, overpressured stages of
evolution, the source may have sufficient energy to increase the
central entropy of the system, whilst at later stages, the source may
only have sufficient energy to control cooling, and no more.  There
are, observationally, some indications that repeated AGN outbursts may
have an effect on the IGM (e.g. \citealt{2005MNRAS.357..279C},
\citealt{2007MNRAS.376..193J}); however, in order to investigate fully
the questions outlined above, systems showing multiple outbursts are
required.  Such systems are relatively rare.  Firstly to detect such
systems, it must be possible to image low surface brightness features
in the X-ray or radio; secondly, and more importantly, it must be
possible to study the gaseous environments of the sources in detail.
Relatively few even of the known multiple-outburst sources meet these
criteria.

The few well-studied systems that show evidence for multiple AGN outbursts
include Centaurus A \citep{2003ApJ...592..129K}, M87
\citep{2000ApJ...543..611O} Perseus~A \citep{2003MNRAS.344L..43F} and
Hydra~A \citep{2007ApJ...659.1153W}.  Cen A does show a strong
shock around the inner radio lobes; however, it is an extremely poor,
very close system ($kT\simeq 0.3$~keV; $D = 3.7$ Mpc), which, whilst
providing an ideal laboratory to study the shock surrounding the new,
young radio source, does not provide an suitable environment for
studying the long term effects of the older outbursts, due to the
paucity of gas in the outer regions and the large angular scale of the
oldest radio structures. M87, Perseus~A and Hydra~A, on the
other hand, are highly complex systems, showing signs of multiple
generations of outbursts, which makes it difficult to disentangle the
effects of successive outbursts. Further, in all these cases, the
youngest components of the radio sources are relatively mature, and
past the shock-heating stage. The only other published example of a
strong shock around a young radio source is NGC~3801
\citep{2007ApJ...660..191C}, which, like Cen A, is a low-temperature
($kT=0.23~\mathrm{keV}$) system in an elliptical galaxy that has
recently undergone a major merger. However, there is little evidence
in this system of multiple AGN outbursts, which are required in order
to investigate the issues raised above.

In this paper we present {\it Chandra} observations of the radio
source B2 0838+32A, otherwise known as 4C\,32.26. This object first
came to our attention as a potential wide-angle tailed radio source
(WAT) during preliminary study of Abell clusters for the work of
\citet{2006MNRAS.368..609J}. Upon closer investigation of the radio
data we found that the source was not a WAT, but rather a restarting
radio source, showing evidence for two distinct epochs of radio
activity, which prompted our {\it Chandra} observations. We begin the
present paper by resolving a long-standing confusion about the
association of this object with an Abell cluster. In the literature,
the host galaxy of the radio source has been presented as the
brightest cluster galaxy (BCG) of Abell~695
\citep[e.g.][]{1978A&AS...34..341F}, but there is an 8-arcmin
separation between the position of the radio galaxy and the catalogued
position of the Abell cluster \citep{1989ApJS...70....1A}. In
Section~\ref{identify} of the present paper we discuss the true
environment of the radio source. We then discuss the details of the
radio observations used in the paper in Section~\ref{radio}. In
Section~\ref{xray} we present our new \Chandra observation and
in Section~\ref{group} we discuss the details of the IGM. Our analysis
of the structures seen in the {\it Chandra} data is presented in
Section~\ref{feature}, the interpretation of our results in terms of
feedback models is given in Section~\ref{consequences} and our
conclusions are presented in Section~\ref{conclusions}.

\section{Identity and redshift of the group hosting B2 0838+32A}
\label{identify}
The radio source B2 0838+32A has traditionally been assumed to reside at
the centre of the galaxy cluster Abell~695 (e.g.
\citealt{1978A&AS...34..341F}), despite the 8~arcmin discrepancy
between the co-ordinates of the radio source (08h41m13.1s, +32d25m00s)
and the galaxy cluster known as Abell~695 (08h41m24.2s, +32d17m16s),
which exceeds the notional 2.5 arcmin uncertainty on the cluster
position \citep{1989ApJS...70....1A}. Additionally, the redshift of
the radio galaxy has almost always been assumed to be the same as that
of the cluster Abell~695, and is cited in the literature as such (e.g.
\citealt{1978AJ.....83..904S}, \citealt{1981ApJS...45..613N},
\citealt{1981SvA....25..647F}, \citealt{1982A&A...108L...7S},
\citealt{1987ApJS...63..543S}, \citealt{1999ApJS..125...35S}). This
has been further reinforced by the presence of a group of galaxies
clustered around B2 0838+32A both spatially and in redshift space (see
Table~\ref{optical:table:positions}).

\begin{table}
\caption{Positions and velocities of the galaxies used to determine
the redshift and velocity dispersion of the Abell~695 group.  All
positions and redshifts are taken from the SDSS Data Release
6.  The final column indicates whether the galaxy is a group member as
classified by {\sc rostat}.}
\label{optical:table:positions}
\begin{tabular}{cccc}\hline
\multicolumn{2}{c}{Galaxy co-ordinates}&$z$&Group Membership\\
$\alpha_{2000}$ &$\delta_{2000}$ &&\\ \hline
08 41 12.79& 32 24 55.20 & 0.0696&\\
08 41 6.22& 32 25 6.79 & 0.0657&\\
08 41 21.06& 32 24 42.74 & 0.0696&\\
08 40 53.81& 32 23 46.17 & 0.0670&Y\\
08 41 8.89& 32 20 38.34 & 0.0683&Y\\
08 41 27.35& 32 28 33.01 & 0.0646&\\
08 41 22.95& 32 28 34.81 & 0.0665&\\
08 41 29.33& 32 24 54.10 & 0.0656&\\
08 40 49.25& 32 25 1.01 & 0.0677&Y\\
08 41 27.72& 32 19 41.17 & 0.0656&\\
08 41 5.64& 32 31 22.76 & 0.0677&\\
08 40 40.16& 32 22 54.08 & 0.0684&Y\\
08 40 12.54& 32 14 22.61 & 0.0683&Y\\
08 40 37.84& 32 12 56.59 & 0.0672&Y\\
08 41 50.17& 32 26 23.48 & 0.0542&\\
08 42 1.38& 32 30 28.25 & 0.0528&\\
08 41 52.92& 32 36 5.98 & 0.0660&\\
08 42 13.92& 32 33 49.72 & 0.0532&\\
08 41 53.93& 32 17 57.45 & 0.0695&Y\\
08 41 53.57& 32 19 59.88 & 0.0671&Y\\
08 40 13.80& 32 27 52.94 & 0.0682&Y\\
08 40 38.97& 32 30 27.34 & 0.0517&\\
08 40 59.09& 32 38 12.39 & 0.0682&Y\\
08 40 54.81& 32 44 28.21 & 0.0684&Y\\
08 42 13.24& 32 17 11.07 & 0.0677&Y\\
08 40 29.24& 32 38 34.96 & 0.0686&Y\\ \hline

\end{tabular}
\end{table}

To determine whether the group associated with the radio galaxy B2
0838+32A and the galaxy cluster Abell~695 are the same entity, we
obtained the SDSS DR 6\footnote{http://cas.sdss.org/astro/en/}
positions and redshifts for all galaxies in a 20~arcmin radius of the
catalogued central position of Abell~695, to ensure that galaxies from
both Abell~695 and those surrounding the radio galaxy were included.
However, examining the SDSS data, whilst it is clear that there is a
group of galaxies around B2 0838+32A, this group does not fit with the
documented Abell richness classification, in that the documented Abell
richness class is 1, indicating that the cluster should contain
between 50-79 galaxies in the magnitude range $m_3$ to $m_3 + 2.0$,
where $m_3$ is the magnitude of the third brightest galaxy, within
1.7~arcmin of the cluster centre \citep{1958ApJS....3..211A}.  As we
have only 26 galaxies in total within 10~arcminutes of the radio
source, and the radio source is offset by 8 arcmin from the central
position of the Abell cluster, we propose instead that the group of
galaxies centred around B2 0838+32A is a real foreground group, whilst
the original system identified by Abell is a cluster of galaxies
centred approximately 8~arcmin south of B2 0838+32A at a much higher
redshift.  This has been suggested in the past by
\citet{1982A&A...108L...7S}, who calculated a photometric redshift of
0.152 for the cluster of galaxies at the catalogued position of
Abell~695.

\begin{figure}
\scalebox{.4}{\includegraphics{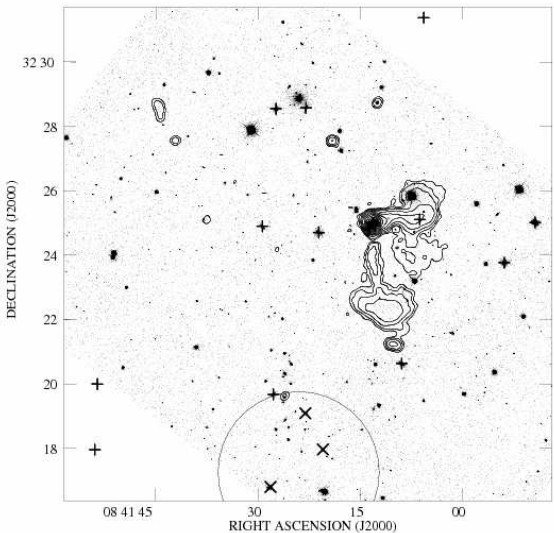}}
\caption{Optical SDSS image of the region overlaid with radio contours
of the radio source B2 0838+32A, together with galaxies at the group
redshift marked with crosses, galaxies in the range $z=0.15-0.18$ in
the cluster region marked with 'x', and the nominal Abell cluster
position marked with a large circle.  It can be seen that there are
two distinct groupings of objects -- a nearby group centred around the
radio galaxy and a more distant group/cluster at higher redshift.}
\label{sdss}
\end{figure}

In Figure~\ref{sdss}, we show the optical image together with
positions (with the redshift range indicated by the different symbols) for
all galaxies within a 20~arcmin radius of the nominal centre of
Abell~695.  It is immediately clear that there are in fact two
overdensities of galaxies, which whilst appearing to be
adjacent/overlapping in projection, are greatly separated in redshift
space.  Using {\sc rostat} \citep{1990AJ....100...32B}, a $3-\sigma$
clipping algorithm, we were able to determine average redshifts and
velocity dispersions for the closer system, but there were
insufficient redshift measurements to allow us to do the same for the
more distant system, which we estimate, using the available SDSS
redshifts for this system, to be in the redshift range 0.15-0.18,
consistent with the estimate of \citet{1982A&A...108L...7S}.

For the closer system, we used SDSS redshifts of objects within a
radius of 10~arcmin of the radio galaxy in the redshift range
$0.06\leq z \leq 0.078$ (the redshift of the radio galaxy is 0.068), a
total of 26 objects in all (all of which fall within the 20~arcmin
radius initially investigated), as detailed in
Table~\ref{optical:table:positions}. We then applied the {\sc rostat}
algorithm to this list of redshifts in order to determine a mean
redshift and velocity dispersion. The algorithm rejected 14 of the
objects on the basis of $3-\sigma$ clipping, leaving us with the 12
objects marked. We obtained $z_{ave}=0.0682$ and
$\sigma_{grp}=300^{+100}_{-50}~\mathrm{km\ s^{-1}}$ for this
foreground system\footnote{Our value of $\sigma$ is consistent within
the joint errors with the somewhat higher value, also determined from
SDSS data, that is tabulated by \protect{\citet{2007A&A...471...17A}}.
The data they present are insufficient to show which galaxies were
used in their analysis and so a more detailed comparison of their
results with ours is not possible.}, which henceforth, for simplicity,
we will call `the 0838+32A group', in contrast to the more distant
Abell~695 cluster. For the remainder of this paper, we take the
redshift of the 0838+32A group, and the radio galaxy it hosts, to be
0.068, which corresponds to a luminosity distance of 300~Mpc in
standard $\Lambda$CDM cosmology, $H_0=71~\mathrm{km\ s^{-1}\
Mpc^{-1}}$; in this cosmology 1~arcsec corresponds to 1.29~kpc.

\section{Radio observations and spectra}
\label{radio}
\begin{figure}
\scalebox{.35}{\includegraphics{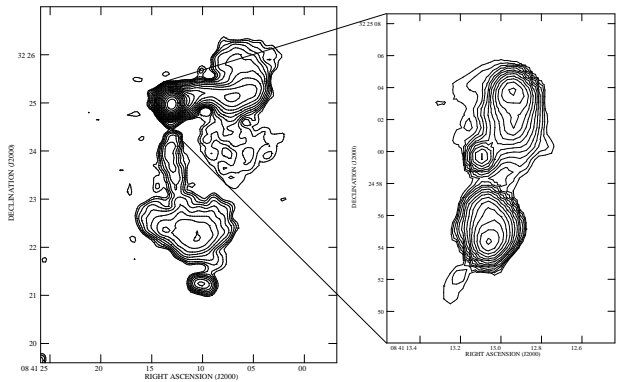}}
\caption{1.4-GHz VLA image of the radio source (main image) of
resolution 14.25 $\times$ 13.02~arcsec, showing the large-scale radio
lobes; the inner lobes appear unresolved in this image.  The inset image is a 1.09
$\times$ 0.98~arsec, 4.9-GHz image that shows the inner core
and lobes.  Contours are in $\sqrt{2}$ steps with the
lowest contour at 0.31~$\mathrm{mJy\ beam^{-1}}$ for both images.}
\label{radio:high}
\end{figure}

We reduced archival 1.4 and 4.8~GHz VLA observations (details given in
Table~\ref{radio:obs}) of the object B2 0838+32A in {\sc aips} in the
usual way.  The low resolution 1.4~GHz map shows a compact central
source embedded in large-scale diffuse emission.  The 4.9-GHz images
show the large-scale structure seen in the 1.4-GHz map and also
resolve the compact inner source as shown in Fig~\ref{radio:high}
(inset).  The bright, compact double lobes of the inner source, with a
total physical scale of 16 kpc, are characteristic of a young radio
galaxy, but the outer lobes, which are a factor $\sim 3$ times fainter
at 1.4 GHz and each extend $\sim 250$ kpc from the nucleus, are
unusual; they both appear to show a jet leading into a broad, faint
lobe, but the `jets' are at 90$^\circ$ to each other in projection,
and show no apparent connection to the inner lobes, a situation
reminiscent of the large-scale `jet' between the inner lobes and north
middle lobe in Cen A \citep{1999MNRAS.307..750M}. As the 4.7-GHz
D-configuration data have similar $uv$ plane coverage to the 1.4-GHz
C-configuration, we were able to measure spectral indices for various
components of the source (for this comparison we discarded the short
baselines that are present in the 1.4-GHz but not the 4.7-GHz dataset
to reduce bias). We found that the inner lobes have
$\alpha_{1.4}^{4.7} = 0.65$ while the outer lobes both have
$\alpha_{1.4}^{4.7} = 1.0$: qualitatively this is consistent with the
idea that the outer lobes are relics of an earlier outburst. The faint
low-surface-brightness southern region of the W lobe is not detected
in the 4.7-GHz data and may be presumed to be steeper-spectrum still.
The large-scale `jets' have intermediate spectral indices.

Fitting \citet{1973A&A....26..423J} spectral aging models to the
lobes assuming an original spectral index of $\alpha =0.55$ (from
comparison with measurements from currently active FRI lobes), we
found that the maximum spectral age of the lobes \citep{Leahy1991} is
around $5 \times 10^7$ years. Losses to inverse-Compton scattering
dominate for plausible (equipartition or sub-equipartition) magnetic
field strengths. As usual, spectral aging timescales of this kind
assume passive evolution of the electron population after the original
population has been set up; thus it is probably best to interpret this
age as a lower limit on the age of the lobe, assuming that since the
lobes are likely to be in approximate pressure balance with the IGM,
there has been little expansion since the energy supply was
discontinued.

The minimum pressure (assuming a fully tangled magnetic field so that
$P=(U_e+U_B)/3$) in the large-scale lobes is of the order of $1.2
\times 10^{-14}$ Pa, while in the inner lobes it is $6\times10^{-12}$ Pa.

\begin{table}
\caption{Details of the VLA radio datasets used in this work.}
\label{radio:obs}
\begin{tabular}{lllll}\hline
Obs. ID&Array &Frequency & Observation & Observation\\
&Configuration & (GHz)& date & duration (mins)\\ \hline
AM364&B & 4.9 & 1993-03-28 &56\\
AV265&D & 4.7 &2003-04-11&250\\
AV265&C & 1.4 &2004-04-10&180\\
\hline
\end{tabular}
\end{table}

\label{radio:analysis}

\section{X-ray Observations}
\label{xray}

The galaxy group was observed with {\it Chandra} in two separate
observations (see Table.~\ref{xray:obs} for details) and processed
following the \CIAO online threads, applying the latest calibration
products, removing bad pixels and excluding periods of high
background. After data cleaning, we were left with good time intervals
of 59.7 and 21.6~ks giving a total clean exposure time of 81.3~ks. A
merged data file was created for spatial analysis, but following \CIAO
recommendations, the spectral analysis was carried out by extracting
spectra separately from each events file and and fitting models to
them jointly. We use {\sc xspec}~ 12 for spectral fitting.

\begin{table}
\caption{Details of the \Chandra X-ray observations used in this
work.}
\label{xray:obs}
\begin{tabular}{llll}
\hline Observation &Observation & Unfiltered exposure &Filtered
exposure\\
ID & date & time (ks) & time (ks) \\ \hline
7917&2006-12-27&61&59.7\\
8500&2006-12-30&21.8&21.6\\ \hline
\end{tabular}
\end{table}

\section{X-ray Analysis}
\label{group}
\subsection{Spatial Analysis}
\label{group:sb}

\begin{figure}
\scalebox{.4}{\includegraphics{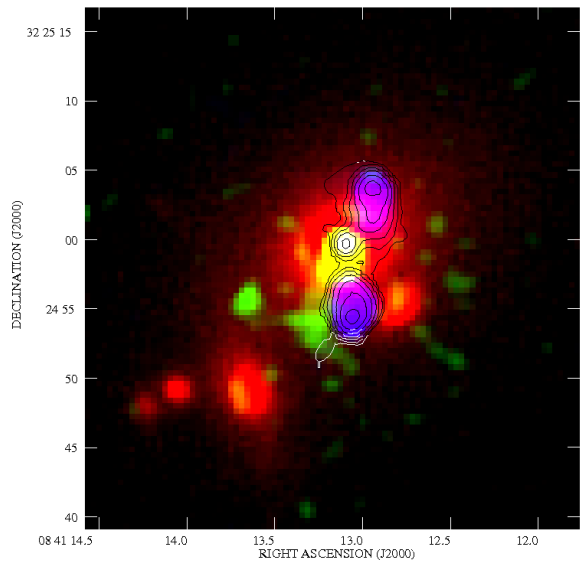}}
\caption{3 colour image incorporating the SDSS R-band data (red
channel), \Chandra X-ray data (green channel), and VLA 4.8~GHz radio
data (blue channel), overlaid with the VLA A-Array 4.8~GHz radio contours to
highlight the radio lobes. It can be seen that the radio lobe is
contained entirely within the BGG of the group and that the X-ray
emission is enhanced in the SE edge of the lobe, but from the SDSS
image, it is clear that the X-rays are being emitted from the region
in between the main galaxy and its smaller companion to the SE.  The
X-ray data are binned in standard {\it Chandra} pixels (0.492 arcsec) and smoothed with a Gaussian of FWHM of
0.5~arcsec.  The radio contours are at $0.25\times\left(\sqrt{2},\
2\sqrt{2},...\right)~\mathrm{mJy\ beam^{-1}}$}
\label{shock:overlay:fig}
\end{figure}

\begin{figure}
\epsfxsize 8.5cm
\epsfbox{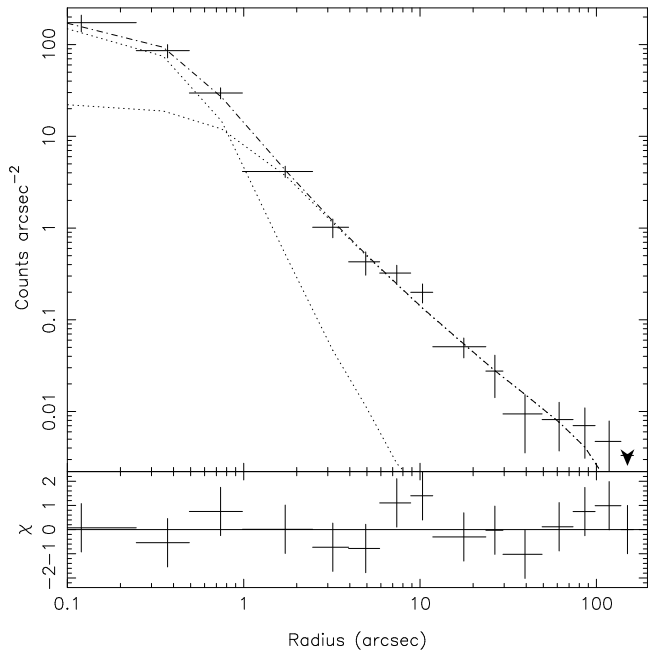}
\caption{The surface brightness profile of the A695 group and central
  galaxy. The dotted lines show the two components of the fitted
  model, consisting of a point source and the double $\beta$ model
  described in the text, both convolved with the {\it Chandra} PSF,
  and the dashed line shows their sum. The residuals (data minus model)
  in terms of the contribution to $\chi^2$ are also plotted.}
\label{group:sb:fig}
\end{figure}

\begin{figure*}
\epsfxsize 16cm
\epsfbox{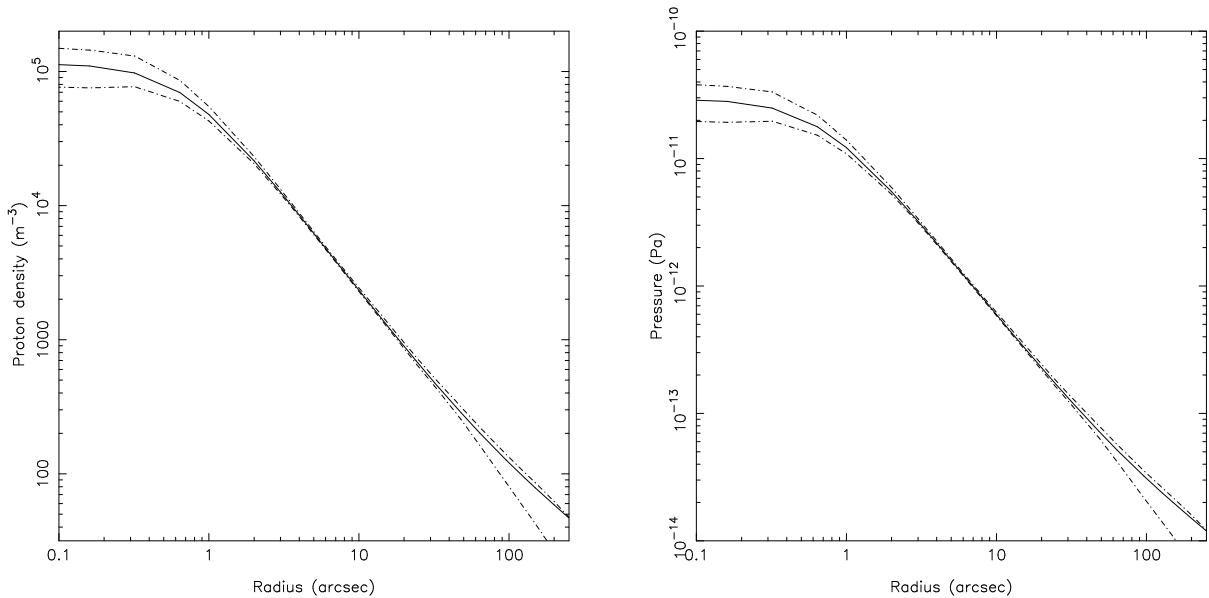}
\caption{The inferred density and pressure profiles for the group
  derived from fitting of the double $\beta$ model described in the
  text. The solid line shows the Bayesian estimate of each quantity,
  and the dashed-dotted lines show the 68\% confidence intervals (formally
  credible intervals) around the estimate. A temperature of 0.7 keV is
  assumed for the gas and the errors on the pressure profile do not
  take into account the uncertainty on this value of temperature.}
\label{group:dp:fig}
\end{figure*}

At first glance, the 0.5-5.0~keV X-ray image shows the AGN and some small-scale
diffuse emission, with structure surrounding the south-eastern edge of
the southern radio lobe (see Fig.~\ref{shock:overlay:fig}), but very
little large-scale emission can be detected by eye. To test for the
presence of any large scale emission, we created the surface
brightness profile shown in Fig.~\ref{group:sb:fig}. This profile
seems to show structure on several scales: the large-scale emission is
not well fitted with a single $\beta$ model. We therefore fitted it
with a projected double $\beta$-model as described by
\citet{2008arXiv0802.4297C}, using the Markov-Chain Monte Carlo
parameter space searching algorithm described in that paper with
uniform priors on $\beta$ values and Jefferys priors on scale
parameters. The best-fitting (maximum-likelihood) model is plotted
with the data in Fig.~\ref{group:sb:fig}. The parameters of the
larger-scale component of the double $\beta$ model are not well
constrained (unsurprisingly, as it is represented by only the few
outer bins of our surface brightness profile) but the data do allow
us to constrain the density and thus the pressure (assuming a
temperature, see Section~\ref{group:spectral}) of thermal material on
the scales covered by the radial profile (Fig.\ \ref{group:dp:fig}).

In order to ascertain how typical the IGM is (e.g. to be able to make
statements about any heating of the IGM by the radio source), we
scaled the surface brightness profile radially and vertically by
$R_{500}$ and compared the profile of the Abell~695 group with that of
other similarly scaled nearby groups taken from
\citet{2007MNRAS.376..193J}.  To determine $R_{500}$ for the Abell~695
group, we used a temperature of 0.7~keV (see
Section~\ref{group:spectral}) and the scaling relation
$R_{500}=391\times T^{0.63}$ from \citet{2005MNRAS.363..675W}, which
we determine to be 310~kpc.  We find that the Abell~695 group is not
significantly different to other local groups, and that the lack of
easily detectable emission is due to the {\sc acis-s} background.

\subsection{Temperature of the group}
\label{group:spectral}

\begin{figure}
\scalebox{.3}{\includegraphics[angle=270]{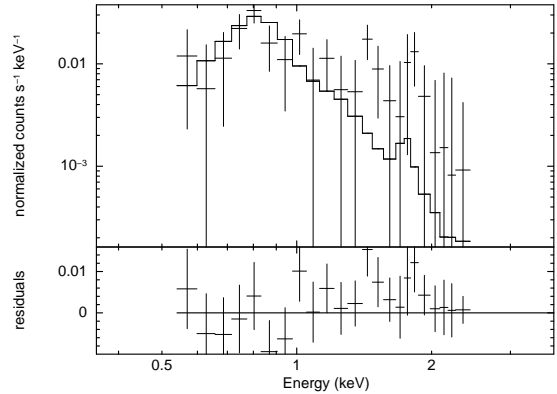}}
\caption{Spectrum of the group, excluding the shock and BGG extracted
from the 7919 observation only and fitted in the energy range
0.5--2.5~keV with an absorbed {\sc apec} model.  We find a temperature
of $0.7\pm 0.1$~keV for the group from this spectrum}
\label{group:fig:spec}
\end{figure}

We extracted spectra from both events files in the same region as used
for the surface brightness analysis in Section~\ref{group:sb}, with a
local background (for each file) for background subtraction, and
excluded a region 7~arcsec (9~kpc) in radius to mask out the
contribution from the brightest group galaxy (BGG) and extended
small-scale structure.  Each spectrum was binned such that each bin
contained 200 counts after background subtraction.  Examining the
spectra from each observation in turn, and considering the faintness
of the source, we found that the emission above 2.5~keV in observation
7919 was dominated by background, whilst the emission at all energies
in observation 8500 was dominated by the background.  Thus, for the
spectral fitting of the group, we use only observation 7919 in the
energy range 0.5--2.5~keV.  We fitted the spectrum with an absorbed
{\sc apec} model with abundance fixed to 0.3\Zsol.  This fit gave a
global group temperature of $0.7\pm0.1$~keV with reduced chi-squared,
\rchisq=0.9 for 21 d.o.f.  We give details of this fit in
Table~\ref{group:par:tab}, and show the fitted spectrum in
Fig~\ref{group:fig:spec}.

To check the veracity of this result, since the spectrum is rather
noisy, we checked the temperature in two ways.  First, we extracted a
spectrum from the entire region, including the BGG (but not the
central AGN, which we exclude using a region of radius 2.5~arcsec) and
surroundings (to provide a better SNR), and a separate spectrum from
simply the BGG (but again with the AGN excluded) and surroundings. We
fitted the BGG spectrum with two absorbed {\sc apec} models (one each
for the galaxy and the extended structure), as detailed in
Table~\ref{gal:par:tab}, and then used these fitted parameters as a
model for the galaxy and shock in the group spectrum.  Modeling the
galaxy and shock and fitting a further {\sc apec} model to the global
spectrum resulted in a temperature of $0.7\pm0.2$~keV with \rchisq=0.8
for 21 d.o.f., in agreement with the temperature found when excluding
the galaxy emission.

We further estimate the temperature based on the $\sigma -T_X$ relation
of \citet{OP2004}.  From the SDSS data in Section~\ref{identify}, we
calculated that $\sigma_{grp}=300^{+100}_{-50}~\mathrm{km\ s^{-1}}$, and
using this velocity dispersion, we find that the predicted temperature
is $0.5^{+0.5}_{-0.2}~\mathrm{keV}$, in agreement with our spectral
measurements of the group temperature.

\begin{figure}
\scalebox{.5}{\includegraphics{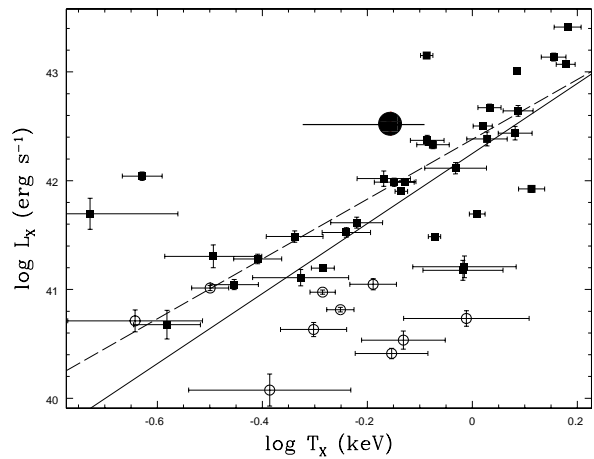}}
\caption{$L_{\mathrm{X}}:T_{\mathrm{X}}$ relation taken from Osmond \&
Ponman (2004); the solid line represents the fit to their entire
sample and the dashed line the fit to only those systems which show
large-scale extended emission.  The open points represent those
sources without large scale extended emission, whilst the closed
points show those systems with extended halos.  The X-ray properties
of the B2~0838+32A group are indicated by the large black point.}
\label{group:fig:lt}
\end{figure}

Thus, taking these multiple temperature estimates into account, we
adopt a temperature of 0.7~keV for the group.  A summary of our
temperature estimates are tabulated in Table~\ref{gal:par:temps}.  We
also plot the group on the $L_X:T_X$ relation of \citet{OP2004}, as
shown in Fig.~\ref{group:fig:lt} using an X-ray temperature of $0.7\pm
0.1$~keV and the unabsorbed X-ray luminosity given in
Table~\ref{group:par:tab}, and the group does not show any significant
deviation from the $L_X:T_X$ relation.

\begin{table}
\caption{Details of the spectral model fitted to the group.}
\label{group:par:tab}
\begin{tabular}{lll}\hline
Component& Parameter& Value \\ \hline  
wabs& $N_{\mathrm{H}}$& $3.96\times 10^{20}~\mathrm{cm^{-2}}$ \\ \hline
{\sc apec}&$kT$&0.7$\pm$0.1~keV\\
               &N&$\left(5.4\pm 0.9\right)\times 10^{-5}$\\ \hline
&$L_{\mathrm{X}}$ (unabsorbed)&$\left(3.2\pm0.2\right)\times
10^{42}~\mathrm{erg\ s^{-1}}$\\
&  \chisq (d.o.f.) & 8.3 (9)\\ 
\hline
\end{tabular}
\end{table}

\begin{table}
\caption{Details of the spectral model fitted to the galaxy and surrounding structure.}
\label{gal:par:tab}
\begin{tabular}{lll}\hline
Component& Parameter& Value \\ \hline 
wabs& $N_{\mathrm{H}}$& $3.96\times 10^{20}~\mathrm{cm^{-2}}$ \\ \hline
{\sc apec}&$kT$&$0.2^{+0.4}_{-0.1}$~keV\\
               &N&$\left(1.5\pm 0.5\right)\times 10^{-5}$\\ \hline
{\sc apec} (shock)&$kT$     & $1.9^{+0.8}_{-0.3}~\mathrm{keV}$\\
          & $N$ & $\left(1.1\pm 0.2\right)\times10^{-4}$\\ \hline
&\chisq (d.o.f.) & 18.9 (21)\\
\hline
\end{tabular}
\end{table}

\begin{table}
\caption{A summary of our various temperature estimates}
\label{gal:par:temps}
\begin{tabular}{ll}\hline
Method & Temperature\\ \hline
Direct fitting to group spectrum & $0.7\pm 0.1$~keV\\
Modeling shock and galaxy &$0.7\pm 0.2$~keV\\
component and fitting group spectrum&\\
Using $\sigma -T_X$ relation of \citet{OP2004}&
$0.5^{+0.5}_{-0.2}$~keV\\
\hline
\end{tabular}
\end{table}

\subsection{Spectrum of the AGN}
\label{agnspec}
In order to be able to compare the nuclear properties of the radio
galaxy with other sources that are known to drive strong shocks
through the IGM, we extracted a spectrum of the AGN, centred on the
radio core and extending out to 2.5~arcsec.  For the background, a
similar sized region was used, positioned so as to avoid the region of
hot gas to the SE.  We fitted the spectrum in the energy range
0.4-7.0~keV with an absorbed power law model, and initially fixed the
absorption to the Galactic value, resulting in a power-law with photon
index, $\Gamma = 1.7 \pm 0.2$, with \rchisq=1.2 (for 6
d.o.f.). Fitting a model similar to that used for Cen.~A provides a
worse fit (\rchisq=2.2 for 3 d.o.f), indicating that the nuclear
spectrum for this source requires a simple power law only.  We find
that the 1-keV nuclear X-ray flux density from the power law is $2.1\pm 0.3$~nJy.
Comparing the X-ray and 5-GHz nuclear fluxes and luminosities (see
Section~\ref{radio}) with the analysis of \citet{2006ApJ...642...96E},
we find that the core properties of B2~0838+32A are entirely consistent with
both the radio and X-ray components being emission from the unresolved
bases of jets, and could also indicate the presence of a radiatively
inefficient accretion flow \citep[e.g.][]{2004ApJ...617..915D}, and a
complete absence of a torus.

\section{The feature to the SE of the radio lobe}
\label{feature}
As can be seen from Fig.~\ref{shock:overlay:fig}, there is a distinct
feature in the X-ray data corresponding to the Southern inner radio
lobe.  From its location and shape, we suggest that this could be part
of a shock generated by an overpressured expanding radio lobe as seen
in systems such as Cen A \citep{2003ApJ...592..129K} and NGC~3801
\citep{2007ApJ...660..191C}.  From Fig.~\ref{shock:overlay:fig}, it is clear that whilst this
feature is only clearly visible around the SE edge of the radio lobe,
it does outline the edge of the radio lobe clearly, and the emission
is not coming from either the main galaxy or the merging galaxy to the
SE. From a comparison with Cen A, where a one-sided shock is also
seen, with the shock being much brighter around the S lobe than the N
lobe, it seems plausible that the reason why the feature is visible
here is that galaxy-galaxy interactions enhance the gas density in the
region.   For the following subsection, we assume, from the
morphology of the feature, that it is a shock generated by the radio
galaxy, and make predictions about the state of the unshocked IGM
that we then test against observations.
\subsection{Shock Analysis}
\label{shock}
In order to determine if the physical parameters of the galaxy and
`shock' are consistent with that of a region containing hot shocked
gas and cooler unshocked gas, we firstly extracted a spectrum of the
central 10~arcsec, containing both the galaxy and the feature in order
to ensure an adequate signal-to-noise ratio to give a well-constrained
fit.  We then fitted this spectrum with an absorbed double {\sc apec}
model in order to constrain the multiple temperatures expected in
a shock model; the results of this fit are given in
Table~\ref{gal:par:tab}.  We find a temperature for the hot component
of the {\sc apec} model of $\left(1.9^{+0.8}_{-0.3}\right)$~keV,
together with the other parameters as detailed in
Table~\ref{gal:par:tab}.  \textbf{ Extracting a spectrum of a region
containing the shock only results in a spectrum with a significantly
worse signal-to-noise ratio, which, whilst still allowing a
temperature to be fitted to the shock, results in a poorly constrained
fit.  The temperature found in this case is consistent with that found
from the multi-temperature model, but very poorly constrained.  Thus,
we use the temperature from the multi-temperature fit in this
analysis.}  Further, since the galaxy temperature is consistent with
the group temperature found in Section~\ref{group:spectral} at
the $1-\sigma$ level, we take the temperature of the group,
$\left(0.7\pm 0.1\right)$~keV to be the temperature of the unshocked
gas.

Assuming that the standard Rankine-Hugoniot conditions hold, the Mach
number, $\mathcal{M}$ of the shock is given by:
\begin{equation}
\frac{T_2}{T_1}=\frac{\left[2\Gamma
\mathcal{M}^2+\left(1-\Gamma\right)\right]\left[\Gamma-1+\frac{2}{\mathcal{M}^2}\right]}{\left(\Gamma
+1\right)^2},
\label{eqn:mach}
\end{equation} where $T_2$ is the temperature of the shocked gas,
$1.9^{+0.8}_{-0.3}$~keV, $T_1$ is the downstream (unshocked) gas
temperature, taken to be the group temperature, $0.7\pm 0.1$~keV, and $\Gamma$
is the adiabatic index, taken here to be 5/3 for a hot,
non-relativistic plasma.  This gives $\mathcal{M}=2.4^{+1.0}_{-0.5}$.

We can then calculate the expected density contrast using:
\begin{equation} \frac{\rho_2}{\rho_1}=\frac{\Gamma
+1}{\Gamma-1+\frac{2}{\mathcal{M}^2}}\label{eqn:machdensity}.\end{equation} This
gives an expected density contrast of $2.6^{+0.6}_{-0.4}$ between the
shocked and unshocked gas.  The pressure contrast is similarly given
by
\begin{equation}\frac{P_2}{P_1}=\frac{2\Gamma\mathcal{M}^2+\left(1-\Gamma\right)
}{\Gamma+1},\label{eqn:machpressure}\end{equation} and found to be
$7^{+7}_{-3}$.

Using the {\sc apec} normalisation $N_S$ for the higher temperature
{\sc apec} model, given in Table~\ref{gal:par:tab}, we can calculate
the density of the shocked gas component since \begin{equation}n_e =
\left\{\frac{4\pi\left[D_A\left(1+z\right)\right]^2 N_{S}\times
1.18}{1\times10^{-14}V_S}\right\}^{\frac{1}{2}} \mathrm{cm^{-3}},
\label{densityeq} \end{equation} where $D_A$ is the angular size
distance to the source, and $V_S$ is the volume of the shocked gas in
cm$^3$.  It is clear that the measured density will be strongly
dependent on $V_S$, but the thickness of the shock cannot be
determined accurately since the shock is not fully resolved in our
data. Comparisons with Cen~A suggest that the shock thickness should
be of the order 200~pc, whilst direct measurement of the shock from
the data suggests a thickness of the order of 2~kpc, although this is
an upper limit, as we have not taken the \Chandra
point-spread-function (PSF) into account.  We thus use these two
values as limits in calculating the density of the shocked gas.  We
model the shock as half a hemispherical shell of inner radius
corresponding to the radius of the radio lobe, measured in {\sc aips}
to be 3.4~arcsec (4.3~kpc), and outer radius as determined by the
shock thickness we use.  Using these densities, we can then calculate
the pressure of the shocked gas using
\begin{equation}P=2.54\times
10^{-9}n_ekT,\label{eqn:pressure}\end{equation} where $n_e$ is the
electron density of the shocked gas, $kT$ is the temperature and $P$
is in Pa.  We show our calculated volumes, densities and pressures
($n_2$ and $P_2$ in Eqns.~\ref{eqn:machdensity} and
\ref{eqn:machpressure} respectively) in Table~\ref{shock:par:table},
together with predicted values of the corresponding unshocked
quantities ($n_1$ and $P_1$).

\begin{table*}
\caption{Estimated parameters for the shocked gas together with
predicted values for the unshocked gas, assuming that standard
Rankine-Hugoniot conditions apply and calculated using two different
values of shock thickness ($\Delta R$)-- 0.2~kpc (by comparison with
Cen A) and 2~kpc (from direct measurement).}
\label{shock:par:table}
\begin{tabular}{cccccc}
\hline
$\Delta R$ & $V_S$ & $\rho_2$ & $P_2$ & Predicted $\rho_1$ & Predicted
$P_1$ \\
(kpc) & ($\mathrm{cm^3}$) & ($\mathrm{cm^{-3}}$) & ($10^{-10}$Pa)
&($\mathrm{cm^{-3}}$) & ($10^{-10}$Pa)\\ \hline
0.2&$3.6\times 10^{65}$&$0.20\pm 0.01$&$9.5^{+2}_{-0.8}$&0.06--0.09&0.62--2.9\\
2 & $5.3\times 10^{66}$&$0.052\pm 0.002$&$2.2^{+0.5}_{-0.2}$&0.02--0.03&0.1--0.7\\
\hline
\end{tabular}
\end{table*}
 
\subsection{Testing predictions from the shock analysis}
\label{comparisions}

If the feature seen in Fig.~\ref{shock:overlay:fig} is indeed a shock,
as indicated by the significantly higher temperature of the region
compared to its surroundings, then the predictions detailed in
Table~\ref{shock:par:table} should agree with the measured parameters
of the unshocked gas. To this end, we use the density profile
calculated in Section~\ref{group:sb} to estimate the density of the
gas that would have been swept up by the inner lobe and subsequently
shocked.  The gas would have been swept up from the central 3~arcsec
(i.e. from within the radius of the lobe), which from Fig.~\ref{group:dp:fig},
is in the range $0.012-0.15~\mathrm{cm^{-3}}$, consistent with our
predictions in Table~\ref{shock:par:table}, for both values of shock
thickness that we use.
  
Furthermore, we calculate from the radio data (see
Section~\ref{radio}) that the minimum
internal pressure of the radio lobe is $P_{min}=6\times10^{-12}$~Pa,
and that the central pressure of the unshocked IGM in the same region
as above is in the range $0.34-2.0\times 10^{-11}$~Pa.  However, as
the radio lobe and shock must be in pressure balance (i.e. the
pressure of the radio lobe should be $2-12\times 10^{-10}$~Pa), this implies
that the true pressure of the radio lobe must be approximately 10-350
times greater than the minimum pressure calculated.  This is consistent with what is
seen in both NGC~3801 and Cen A, where the minimum pressure
calculated directly from the radio lobes is far less than the
shock pressure, so that the lobes must be some way from the minimum
pressure condition.

Additionally, we can investigate the ways in which shock formation may
be suppressed to the SW of the S radio lobe and over the whole of the
N radio lobe. The lobe and the visible shock should be in pressure
balance; this pressure must be constant throughout the radio lobe, and
is what drives the shock.  Further assuming that the N radio lobe has
a similar pressure to the S radio lobe, there are then two ways in
which visible shock formation can be suppressed:
\begin{enumerate}
\item A shock could be formed, but could just be too faint to detect in
this observation.
\item The external pressure and/or density could be increased to such
a value that the expansion is no longer supersonic in these regions,
but rather subsonic, so that no shock is formed.
\end{enumerate}

In the first case if the density decreases, then so will the
external pressure and in turn, the Mach number will also increase,
implying a stronger shock.  However, in the limit of a strong shock,
the density contrast tends to 4. Using this fact, we can place limits on the
density around the SW of the S lobe and the N lobe from the X-ray
data.

From the data, we find that after background subtraction, there are
approximately 72 counts in the shocked region.  Placing similarly
sized regions to the SW, NW and NE of the radio lobes, we find upper
limits on the number of counts in what could be regions of shocked
gas, assuming Poisson statistics, and show these limits in
Table~\ref{table:nondet:limits}.  Then, assuming the same conversion
between counts and density in the SW, NW and NE regions as in the
visibly shocked region, i.e. the same temperature, we can calculate
upper limits on the density as the number of counts, $N_{c} \propto
n_e$.  The calculated upper limits on shocked and unshocked density
are given in Table~\ref{table:nondet:limits}.  From the limits calculated,
we estimate that the SE portion of the lobe could be hitting a region
of the IGM which is 3-5 times denser than the surrounding medium.
This increase in density could arise from interactions with the spiral
galaxy to the SE, causing the shock to be visible only where it is.

\begin{table*}
\caption{Estimated upper limits on density in the regions where the
shock is not detected}
\label{table:nondet:limits}
\begin{tabular}{llll|ll} \hline
Region & Counts &\multicolumn{2}{c}{Estimated shocked density ($\mathrm{cm^{-3}}$)}&\multicolumn{2}{c}{Estimated unshocked density (\protect{$\mathrm{cm^{-3}}$})}\\
& &$\Delta R= 0.2$~kpc & $\Delta R= 2$~kpc &$\Delta R= 0.2$~kpc & $\Delta R= 2$~kpc\\ \hline \hline
SE (visible) & 72&0.20&0.052 & 0.06 - 0.09 & 0.02 - 0.03\\ \hline
SW& 13 & 0.085&0.022 & 0.021 & 0.0055\\
NW & 8 & 0.067 &0.017& 0.017 & 0.0043\\
NE & 10& 0.075 & 0.019& 0.019 & 0.0048\\ \hline
\end{tabular}
\end{table*} 

Alternatively it could be the case that the external temperature or
density were high enough such that the expansion no longer produced a
shock.  This would require the ratio between the internal
and external pressures, $P_{\rm int}$ and $P_{\rm ext}$ respectively, to follow
\begin{equation}\frac{P_{\rm int}}{P_{\rm ext}}\ga\Gamma + 1, \end{equation}
implying that if $P_{ext} \sim 0.5P_{int}$, (i.e. if $P_{ext}$ was
increased to the range $1 - 6 \times 10^{-10}$~Pa) then a supersonic
expansion would no longer occur.  In this case, we would require that
either the temperature would be increased (at constant density) to
approximately 1.9~-~11~keV or that the density would increase (at a
constant temperature of 0.7~keV) to $0.056 - 0.33~\mathrm{cm^{-3}}$.
However, at constant temperature, this would increase the emissivity,
making any emission brighter.  The emissivity could be reduced if the
temperature were also correspondingly reduced, but this would have the
effect of making the shock stronger, and thus increasing the
temperature, implying that increasing the external pressure to cause a
sub-sonic expansion is probably not a viable method of suppressing
shock formation in these circumstances.

It seems clear from our analysis of the unshocked gas and the regions
where a shock would be expected to be seen, that the physical
properties of the system are entirely consistent with the scenario
presented of a young, over-pressured radio galaxy, expanding
supersonically into the IGM and generating shocks in the IGM as it
does so.  The sidedness of the shock can be explained by variations in
the IGM density, and the morphology of the elongated structure seen in
the X-ray is consistent with this feature being a shock generated by
the restarting radio galaxy.

\section{The relationship between the host galaxy, the radio properties and consequences for feedback}
\label{consequences}

It is known that other observed young, overpressured radio sources,
Cen A and NGC~3801, exhibit evidence for their host galaxies having
recently undergone mergers with gas-rich galaxies
(\citealt{1998A&ARv...8..237I} and \citealt{2007ApJ...660..191C}).
Furthermore, they exhibit signs of containing dusty tori (high
absorbing columns are required to model the spectra accurately).  This
is more usually associated with powerful, high-excitation radio
galaxies (generally FRIIs) as described in \citet{2006MNRAS.370.1893H}
and \citet{2007MNRAS.376.1849H}, where it is argued that such sources
accrete cold gas in a conventional radiatively
efficient mode (`cold-mode' accretion).  Recent mergers in both
Cen A and NGC~3801 could provide a reservoir of cold gas that could
fuel this accretion mode.

However, whilst the host galaxy of B2 0838+32A shows signs of
interactions with nearby galaxies, there is no evidence for either the
host galaxy having had a recent merger,nor is there evidence for
large quantities of cold gas and dust in the optical images of the
host galaxy, although we do not have deep optical observations.
Further, the AGN spectrum does not need a high absorbing column to
accurately fit the spectrum.  On the contrary, as seen in
Section~\ref{agnspec}, the best-fitting AGN spectrum is a uniform
power law with Galactic absorption, with no evidence for absorption
from a dusty torus and is consistent in all ways with the class of
low-excitation radio galaxies to which most FRI radio galaxies belong,
and which are argued by
\citet{2007MNRAS.376.1849H} to be powered in a radiatively inefficient
manner, termed a radiatively inefficient accretion flow, (RIAF) by
direct accretion of the hot, X-ray-emitting gas.  Thus, it is possible
that the restarting outburst seen in B2 0838+32A is a purely heating
and cooling regulated event, accreting from the hot IGM rather than
from a reservoir of cold gas that has resulted from a recent merger
with a gas-rich galaxy.  It is probable that the previous outburst,
seen at large scales, heated the gas surrounding it whilst active.
This would have caused the gas around the core to expand as energy was
injected into it, until such time that the gas was too tenuous to feed
the AGN, effectively starving the AGN of fuel.  Without excess heating
from an AGN, the IGM was then able to cool radiatively until such a
point that the black hole was able to accrete hot gas once more and
restart the radio galaxy.  

Whilst we cannot definitively rule out the presence of cold gas in the
radio galaxy host, the X-ray AGN spectrum, as discussed above, rules
out absorption from a dusty torus.  Further deep observations in the
optical and H{\sc I} bands would be needed to conclusively determine
that no cold gas was present, but the lack of absorption from a dusty
torus is consistent with sources where it is thought that hot-mode
accretion is the primary fuel of the radio source.

\subsection{Feedback time-scales and dynamics of the radio sources}
\label{feedbacktimescales}

Using the spectral ages of the outer lobes calculated in
Section.~\ref{radio}, together with dynamical arguments, we can
estimate the time-scales on which feedback processes in this system
occur.  From Section~\ref{radio}, the outer lobes are approximately
$\left(4.6 \pm 0.2\right)\times 10^7$~yr old.  Since this assumes the
initial acceleration of a population of relativistic electrons which
are then left to evolve without further re-acceleration, this
time-scale can be taken to be the time when energy injection ceased,
i.e. when the jets turned off, and gives a lower limit on the age of
the lobes.  We can then calculate an upper limit from the dynamics of
the outer lobes.  If we assume that they have expanded in-situ from a
negligibly small radius to their current radius of approximately
65~kpc at the sound speed of the group (though it is more likely that
they have expanded subsonically), then an upper limit on the age of
the lobes is approximately $2\times 10^{8}$~years, which obviously
includes the time during which the jet was active and after the jet
has been switched off.  The large scale lobes suggest that there may
be some moderate motion of the host galaxy with respect to the IGM in
an east-west direction.  This motion could have disrupted the jets via
ram pressure as seen in narrow (and some wide) angle tailed sources.
However, if the galaxy velocity is subsonic, then our energy estimates
will not be significantly affected.  We calculate that a galaxy
velocity $v_{\mathrm{gal}} \sim 3000~\mathrm{km\s^{-1}}$ would be
required for these lobes to be a passive trail left behind by the
galaxy; if this were the case, the energetics would be somewhat
different, but given that this velocity is some 10 times larger than
the velocity dispersion of the group calculated in
Section~\ref{identify}, we do not regard this model as plausible.

Whilst we cannot obtain a spectral age estimate of the inner lobes,
the flatter spectrum suggests that these must neccesarily be younger
than the outer lobes.  Dynamically, we know that the lobes are
expanding supersonically at $2.4^{+1.0}_{-0.5}$ times the local sound
speed, which we calculate to be 330~$\mathrm{km\ s^{-1}}$.  If we
further assume that the initial radius of the lobes is negligible
compared to the current radius, which we take to be 4.3~kpc, then the
inner lobes must be between 3.4 - 6.5~Myr old.

Furthermore, the cooling time of the gas near the centre of the group
is approximately 170~Myr (assuming $kT=0.7$~keV, the density of the
gas is in the range 0.078-0.015~$\mathrm{cm^{-3}}$ and that energy
radiates away at $L_{\mathrm{X}}$ as given in
Table~\ref{group:par:tab}).  This cooling time-scale is consistent
with the available constraints on the ages of the lobe, derived above,
which suggests that cooling gas could be responsible for
retriggering the AGN, and that a feedback driven system, with the
central AGN being responsible for heating its surroundings until it
starves itself of fuel, is plausibly at work in this group.

\subsection{Energetics of the inner lobes}
\label{energy:inner}
In such a case as described above, one would expect that the power
available from Bondi accretion should be able to account for the power
of the inner lobes of the radio galaxy (see for instance,
\citealt{2007MNRAS.376.1849H}, who argue that hot-mode/Bondi accretion
is sufficient to power all objects such as 0838+32A).  Thus, by
estimating the black hole mass ($M_{\mathrm{BH}}$) of the host galaxy,
together with suitable estimates of upper limits to the central
density ($n_0$), an estimate can be made of the Bondi power
($P_{\mathrm{B}}$) available to the radio source. We take the upper
bound of the central density found in Section~\ref{group:sb} as a
constraint on the density of the gas available to the AGN,
i.e. $n_0=0.1~\mathrm{cm^{-3}}$.  We estimate the black hole mass
using the relation between black hole mass and K-band bulge magnitude
derived by \citet{2003ApJ...589L..21M}
\begin{equation}\log_{10}M_{\mathrm{BH}} = 8.21
+1.13\times\left(\log_{10}L_K - 10.9\right). \label{MBH}\end{equation}
Obtaining an absolute K-band luminosity of $L_{\mathrm{K}}=5.1\times
10^{11}~L_{\odot}$ from the 2MASS survey, we find that
$M_{\mathrm{BH}}=1.33\times 10^9$\Msol. The Bondi accretion rate is
given by \begin{equation}\dot{M_{\mathrm{B}}}=\frac{\pi \rho_a G^2
M^2_{\mathrm{BH}}}{c^3_s}, \label{bondiac}\end{equation} where $c_s =
\sqrt{\left(\Gamma kT\right)/\left(\mu m_p\right)}$ is the sound speed
of the hot gas. We find that $\dot{M_{B}} = 0.063$~\Msol~yr$^{-1}$,
and that assuming an efficiency of 0.1, $P_{\mathrm{B}}=3.6\times
10^{37}$~W. To see if hot-mode accretion could account for the power
of the radio source, we can calculate the work done ($W$) on the IGM
by the radio lobe since the internal pressure ($P$) of the radio lobes
is known (see Table~\ref{shock:par:table}), and the volume ($V$) can
be estimated from the high-resolution 4.8~GHz radio maps assuming a
lobe radius of 4.3~kpc and spherical symmetry to be $9.78\times
10^{60}\mathrm{m^3}$.  Further assuming that the plasma is
relativistic, then the work done and the energy stored in the lobes is
given by: 
\begin{equation}W=4PV,\label{4pv}\end{equation} from which we find that
$W=\left(1.1 - 6.7\right)\times 10^{52}$~J.  Furthermore, given that
the Mach number of the visible shock is known, the time-scale for
supersonic expansion must be in the range 3.4 - 6.5~Myr, giving a
mechanical power output for the radio source of $\left( 5.4 -
62\right)\times 10^{37}$~W. This is comparable with $P_{\mathrm{B}}$, given
the uncertainties on both the central density and $M_{\mathrm{BH}}$
used to calculate $\dot{M_{B}}$ and hence $P_{\mathrm{B}}$,
and particularly given that our radial profile analysis could not have
detected an increase in the gas density on scales smaller than a few
kpc, so that the central density we estimate is really a lower limit
on the density available at the Bondi radius.

\subsection{Energetics of the outer lobes}
\label{energetics:outerlobes}
In addition to calculating the work done by the inner lobes, we can
also calculate the energy required to power the large outer lobes as they
gently expand in order to reach pressure equilibrium.  In this case,
if we assume from the presence of relic jets seen at large scales,
that the outer lobes were created in-situ and since the large scale
jets switched off the lobes have been expanding sub-sonically, we can
calculate the work being done on the intra-group medium (IGM) at large
scales and thus compare the relative energy inputs from a radio source
at different stages of evolution.  From the 1.4~GHz radio maps, we
calculate that the NW lobe is 100~arcsec from the centre of the group
and that the S lobe is 150~arcsec from the centre, and from
Fig.~\ref{group:dp:fig}~(right), we find that the pressures at these
radii are $2-3\times 10^{-14}$~Pa and $1-2\times
10^{-14}$respectively.  We further model the lobes as ellipsoids, with
the NW lobe having $a_{\mathrm{NW}}=80$~kpc and
$b_{\mathrm{NW}}=c_{\mathrm{NW}}=60$~kpc, and the S lobe having
$a_{\mathrm{S}}=90$~kpc and $b_{\mathrm{S}}=c_{\mathrm{S}}=60$~kpc.
If we also assume that the lobes' initial volume is negligible
compared to their current volume, the lobes are in pressure balance
with the IGM and that projection does not greatly affect the lobes
(projection effects would make the lobes larger but place them in a
lower-pressure environment), we find that for a direct comparison
with the energetics of the inner lobes using Eqn.~\ref{4pv}, the total
amount of energy supplied to the NW lobe is $8-12\times 10^{51}$~J and
$16-32\times 10^{50}$~J to the S lobe. It is clear that both the inner
lobes, and the older outer lobes are energetically comparable, with
the smaller lobes being slightly more energetic, suggesting that in
both cases, the radio source is able to regulate cooling in the
group atmosphere.

The total $P\mathrm{d}V$ work done by the outer lobes is of the order $2
- 4 \times 10^{51}$~J, and, using the limits on the age of the lobes
given in Section~\ref{feedbacktimescales}, this implies a mean rate of
energy input into the IGM of between $3 \times 10^{35}$ and $3 \times
10^{36}$~W. Within the uncertainties, this is very comparable to the
bolometric X-ray luminosity of the group, $(3.2 \pm 0.2) \times
10^{35}$~W, which implies that the energy supplied by the outer lobes
has been close to sufficient to balance radiative losses over their
lifetime.  However, at this phase in the source's life, any energy
input from the outer lobes will be at large radii, and would not be
expected to prevent the cooling of the inner parts of the hot IGM.

\subsection{Implications for feedback models}
\label{implicationsforfeedback}
As calculated and discussed in the preceding Sections, there is
sufficient energy in both the small-scale and large-scale lobes to
counteract the effects of radiative cooling. The shock currently
being driven by the small-scale lobes will also increase the central
entropy and (since the Bondi accretion rate depends on entropy, $K =
T/n^{2/3}$, as $K^{-3/2}$) will slow or stop accretion on to the
central black hole. However, if the lobes from the previous outburst
were in fact regulating accretion, as required in feedback models,
then the lobes must have grown to quite a large size before accretion
stopped completely; it is not clear in this case how exactly the
energy/entropy would have been provided to the accreting
gas. Moreover, as discussed in Section \ref{energy:inner}, the current,
small-scale outburst does contain enough energy to counteract cooling
immediately, provided that all the energy can be injected directly
where it is needed. As the outburst is still active, it is clear that
there is a time delay between the injection of energy and the
shutting down of accretion. 

This time delay could arise either because it takes some time for the
energy/entropy input from the radio source to diffuse to where it is
needed, or because some time is required to drain the accretion flow
even after the conditions for rapid accretion cease to be met. In
general a characteristic timescale for accretion flow emptying is
$M/{\dot M}$, where $M$ is the mass in the accretion flow and $\dot M$
is the accretion rate. For accretion from the hot phase, we can
estimate that $M > {4\over 3}\pi R_{\rm Bondi}^3 \rho$, and this
combined with equation (8) gives a lower limit for the emptying time
which is dependent only on the temperature and the central black hole
mass. For a system where the temperature near the Bondi radius is 0.2
keV and the black hole has a mass of $10^9 M_\odot$ (see
Section~\ref{energy:inner}) we estimate that this lower limit is $\sim
2 \times 10^6$ years. While this limit is considerably less than the
current source on-time, it is within an order of magnitude, and so it
does not seem implausible that the timescale for the accretion flow to
empty could help to account for the delay in the response of the AGN
to changing external conditions.  Further, any angular momentum
present in the accretion flow will contribute to the mass of any gas
reservoir present, increasing the reservoir emptying time above that
predicted from the Bondi formula.

In addition, we have shown that the central cooling time and the time
since the old source switched off are very similar; thus there is
another time delay in the feedback process, involving a timescale
comparable to the central cooling time. While the time delays
discussed in the preceding paragraph are expected in all
feedback models, the delay involving the cooling time depends on the
detailed microphysics of energy/entropy transfer. However, it is clear
that the relationship between the reservoir emptying/entropy raising
timescale and the cooling timescale in a given system must determine
the `on-time' and duty cycle of a particular radio source.

\section{Summary and conclusions}
\label{conclusions}

We have presented \Chandra observations of the group of galaxies
hosting the radio source B2 0838+32A, and having shown that the host
group is a foreground group at a redshift 0.068, unassociated with the
Abell cluster Abell~695, we have then investigated the interaction
between the radio galaxy and its host group. Our main conclusions can
be summarized as follows:

\begin{enumerate}

\item From the radio morphology and spectra, we propose that radio
galaxy has recently restarted, exhibiting two distinct cycles of AGN
activity; an older, `dead' source that is no longer being fueled
(thought to be between 50 - 200~Myr old), and a pair of smaller,
younger lobes (estimated to be between 3 - 7~Myr old).

\item The younger lobes are over-pressured, and expanding
supersonically and driving a shock into the ISM.  This shock has a
Mach number of $2.4^{+1.0}_{-0.5}$, and we are able to calculate the
density and pressure of the shocked gas.  Further, we make predictions
about the state of the unshocked gas and regions where the shock is
not visible, and find that the predictions are consistent with the
data.

\item As the radio lobe must be in pressure balance with the shock, we
find that the pressure of the radio lobe is approximately 20-100 times
greater than the minimum pressure calculated from radio observations
alone, consistent with what is seen in Cen A and NGC~3801.

\item The one sided-nature of the shock, which is similar to what is seen
in Cen A where the shock is much fainter around the N lobe than the S
lobe, can be explained by a varying density distribution across the
BGG environment, which could be caused by the interacting galaxies in
this system.

\item The older lobes have done sufficient $P\mathrm{d}V$ work on the
IGM over their lifetime to counteract radiative cooling at large
radii.

\item However, in contrast to Cen A and NGC~3801, it appears plausible
that B2 0838+32A is not being fueled from a reservoir of cold gas
arising from a merger, and neither is there any evidence for a dusty
torus to fuel the accretion. Whilst further observations would be
needed to definitively rule out the presence of very cold accreting
gas, the absence of a dusty torus provides evidence for a radiatively
inefficient warm accretion flow, establishing observationally for the
first time that feedback controlled radio galaxy outbursts can give
rise to entropy-changing events such as shocks in the ISM.

\item The energetics of the system imply that there is a time delay in
stopping the accretion (and therefore turning off AGN activity) which
is likely to be related either to the time taken to raise the entropy
of the gas close to the black hole sufficiently to stop accretion, or
to the time taken to empty the gas from the accretion flow. In
addition, based on our estimates of the timescales of the multiple
outbursts in the radio galaxy, we argue that the time delay for the
AGN activity to restart is comparable to the cooling time of the gas
at the centre of the system. We hypothesise that it is the combination
of these two timescales that produces the 'on-time' and duty cycle
times for a given radio source.

\end{enumerate}
\section*{Acknowledgments}
NNJ thanks CNES (the French Space Agency) for funding.  MJH thanks the
Royal Society for a research fellowship.  The authors would also like
to thank Arif Babul for useful discussions regarding the energetics
and timescales and Christian Kaiser for discussion of accretion
timescales.
\bibliographystyle{mn2e} \bibliography{MN-08-0989-MJ-R1-bib}

\begin{thebibliography}{}

\bibitem[\protect\citeauthoryear{{Abell}}{{Abell}}{1958}]{1958ApJS....3..211A}
{Abell} G.~O.,  1958, \apjs, 3, 211

\bibitem[\protect\citeauthoryear{{Abell}, {Corwin} Jr. \& {Olowin}}{{Abell}
  et~al.}{1989}]{1989ApJS...70....1A}
{Abell} G.~O.,  {Corwin} Jr. H.~G.,    {Olowin} R.~P.,  1989, \apjs, 70, 1

\bibitem[\protect\citeauthoryear{{Aguerri}, {S{\'a}nchez-Janssen} \&
  {Mu{\~n}oz-Tu{\~n}{\'o}n}}{{Aguerri} et~al.}{2007}]{2007A&A...471...17A}
{Aguerri} J.~A.~L.,  {S{\'a}nchez-Janssen} R.,    {Mu{\~n}oz-Tu{\~n}{\'o}n} C.,
   2007, \aap, 471, 17

\bibitem[\protect\citeauthoryear{{Beers}, {Flynn} \& {Gebhardt}}{{Beers}
  et~al.}{1990}]{1990AJ....100...32B}
{Beers} T.~C.,  {Flynn} K.,    {Gebhardt} K.,  1990, \aj, 100, 32

\bibitem[\protect\citeauthoryear{{Begelman}}{{Begelman}}{2001}]{2001ASPC..240.%
.363B}
{Begelman} M.~C.,  2001, in {Hibbard} J.~E.,  {Rupen} M.,   {van Gorkom} J.~H.,
   eds, ASP Conf. Ser. 240: Gas and Galaxy Evolution {Impact of Active Galactic
  Nuclei on the Surrounding Medium}.
pp 363--+

\bibitem[\protect\citeauthoryear{{B{\^i}rzan}, {Rafferty}, {McNamara}, {Wise}
  \& {Nulsen}}{{B{\^i}rzan} et~al.}{2004}]{2004ApJ...607..800B}
{B{\^i}rzan} L.,  {Rafferty} D.~A.,  {McNamara} B.~R.,  {Wise} M.~W.,
  {Nulsen} P.~E.~J.,  2004, \apj, 607, 800

\bibitem[\protect\citeauthoryear{{Clarke}, {Sarazin}, {Blanton}, {Neumann} \&
  {Kassim}}{{Clarke} et~al.}{2005}]{2005ApJ...625..748C}
{Clarke} T.~E.,  {Sarazin} C.~L.,  {Blanton} E.~L.,  {Neumann} D.~M.,
  {Kassim} N.~E.,  2005, \apj, 625, 748

\bibitem[\protect\citeauthoryear{{Croston}, {Hardcastle} \&
  {Birkinshaw}}{{Croston} et~al.}{2005}]{2005MNRAS.357..279C}
{Croston} J.~H.,  {Hardcastle} M.~J.,    {Birkinshaw} M.,  2005, \mnras, 357,
  279

\bibitem[\protect\citeauthoryear{{Croston}, {Hardcastle}, {Birkinshaw},
  {Worrall} \& {Laing}}{{Croston} et~al.}{2008}]{2008arXiv0802.4297C}
{Croston} J.~H.,  {Hardcastle} M.~J.,  {Birkinshaw} M.,  {Worrall} D.~M.,
  {Laing} R.~A.,  2008, ArXiv e-prints, 802

\bibitem[\protect\citeauthoryear{{Croston}, {Kraft} \& {Hardcastle}}{{Croston}
  et~al.}{2007}]{2007ApJ...660..191C}
{Croston} J.~H.,  {Kraft} R.~P.,    {Hardcastle} M.~J.,  2007, \apj, 660, 191

\bibitem[\protect\citeauthoryear{{Donato}, {Sambruna} \& {Gliozzi}}{{Donato}
  et~al.}{2004}]{2004ApJ...617..915D}
{Donato} D.,  {Sambruna} R.~M.,    {Gliozzi} M.,  2004, \apj, 617, 915

\bibitem[\protect\citeauthoryear{{Evans}, {Worrall}, {Hardcastle}, {Kraft} \&
  {Birkinshaw}}{{Evans} et~al.}{2006}]{2006ApJ...642...96E}
{Evans} D.~A.,  {Worrall} D.~M.,  {Hardcastle} M.~J.,  {Kraft} R.~P.,
  {Birkinshaw} M.,  2006, \apj, 642, 96

\bibitem[\protect\citeauthoryear{{Fabian}, {Sanders}, {Allen}, {Crawford},
  {Iwasawa}, {Johnstone}, {Schmidt} \& {Taylor}}{{Fabian}
  et~al.}{2003}]{2003MNRAS.344L..43F}
{Fabian} A.~C.,  {Sanders} J.~S.,  {Allen} S.~W.,  {Crawford} C.~S.,  {Iwasawa}
  K.,  {Johnstone} R.~M.,  {Schmidt} R.~W.,    {Taylor} G.~B.,  2003, \mnras,
  344, L43

\bibitem[\protect\citeauthoryear{{Fabian}, {Sanders}, {Taylor}, {Allen},
  {Crawford}, {Johnstone} \& {Iwasawa}}{{Fabian}
  et~al.}{2006}]{2006MNRAS.366..417F}
{Fabian} A.~C.,  {Sanders} J.~S.,  {Taylor} G.~B.,  {Allen} S.~W.,  {Crawford}
  C.~S.,  {Johnstone} R.~M.,    {Iwasawa} K.,  2006, \mnras, 366, 417

\bibitem[\protect\citeauthoryear{{Fanti}, {Gioia}, {Lari} \& {Ulrich}}{{Fanti}
  et~al.}{1978}]{1978A&AS...34..341F}
{Fanti} R.,  {Gioia} I.,  {Lari} C.,    {Ulrich} M.~H.,  1978, \aaps, 34, 341

\bibitem[\protect\citeauthoryear{{Feitsova}}{{Feitsova}}{1981}]{1981SvA....25.%
.647F}
{Feitsova} T.~S.,  1981, Soviet Astronomy, 25, 647

\bibitem[\protect\citeauthoryear{{Hardcastle}, {Evans} \&
  {Croston}}{{Hardcastle} et~al.}{2006}]{2006MNRAS.370.1893H}
{Hardcastle} M.~J.,  {Evans} D.~A.,    {Croston} J.~H.,  2006, \mnras, 370,
  1893

\bibitem[\protect\citeauthoryear{{Hardcastle}, {Evans} \&
  {Croston}}{{Hardcastle} et~al.}{2007}]{2007MNRAS.376.1849H}
{Hardcastle} M.~J.,  {Evans} D.~A.,    {Croston} J.~H.,  2007, \mnras, 376,
  1849

\bibitem[\protect\citeauthoryear{{Heinz}, {Br{\"u}ggen}, {Young} \&
  {Levesque}}{{Heinz} et~al.}{2006}]{2006MNRAS.373L..65H}
{Heinz} S.,  {Br{\"u}ggen} M.,  {Young} A.,    {Levesque} E.,  2006, \mnras,
  373, L65

\bibitem[\protect\citeauthoryear{{Israel}}{{Israel}}{1998}]{1998A&ARv...8..237%
I}
{Israel} F.~P.,  1998, \aapr, 8, 237

\bibitem[\protect\citeauthoryear{{Jaffe} \& {Perola}}{{Jaffe} \&
  {Perola}}{1973}]{1973A&A....26..423J}
{Jaffe} W.~J.,  {Perola} G.~C.,  1973, \aap, 26, 423

\bibitem[\protect\citeauthoryear{{Jetha}, {Hardcastle} \& {Sakelliou}}{{Jetha}
  et~al.}{2006}]{2006MNRAS.368..609J}
{Jetha} N.~N.,  {Hardcastle} M.~J.,    {Sakelliou} I.,  2006, \mnras, 368, 609

\bibitem[\protect\citeauthoryear{{Jetha}, {Ponman}, {Hardcastle} \&
  {Croston}}{{Jetha} et~al.}{2007}]{2007MNRAS.376..193J}
{Jetha} N.~N.,  {Ponman} T.~J.,  {Hardcastle} M.~J.,    {Croston} J.~H.,  2007,
  \mnras, 376, 193

\bibitem[\protect\citeauthoryear{{Kraft}, {V{\'a}zquez}, {Forman}, {Jones},
  {Murray}, {Hardcastle}, {Worrall} \& {Churazov}}{{Kraft}
  et~al.}{2003}]{2003ApJ...592..129K}
{Kraft} R.~P.,  {V{\'a}zquez} S.~E.,  {Forman} W.~R.,  {Jones} C.,  {Murray}
  S.~S.,  {Hardcastle} M.~J.,  {Worrall} D.~M.,    {Churazov} E.,  2003, \apj,
  592, 129

\bibitem[\protect\citeauthoryear{{Leahy}}{{Leahy}}{1991}]{Leahy1991}
{Leahy} J.~P.,  1991, {Interpretation of large scale extragalactic jets}.
Beams and Jets in Astrophysics, pp 100--+

\bibitem[\protect\citeauthoryear{{Marconi} \& {Hunt}}{{Marconi} \&
  {Hunt}}{2003}]{2003ApJ...589L..21M}
{Marconi} A.,  {Hunt} L.~K.,  2003, \apjl, 589, L21

\bibitem[\protect\citeauthoryear{{Morganti}, {Killeen}, {Ekers} \&
  {Oosterloo}}{{Morganti} et~al.}{1999}]{1999MNRAS.307..750M}
{Morganti} R.,  {Killeen} N.~E.~B.,  {Ekers} R.~D.,    {Oosterloo} T.~A.,
  1999, \mnras, 307, 750

\bibitem[\protect\citeauthoryear{{Noonan}}{{Noonan}}{1981}]{1981ApJS...45..613%
N}
{Noonan} T.~W.,  1981, \apjs, 45, 613

\bibitem[\protect\citeauthoryear{{Nulsen}, {McNamara}, {Wise} \&
  {David}}{{Nulsen} et~al.}{2005}]{2005ApJ...628..629N}
{Nulsen} P.~E.~J.,  {McNamara} B.~R.,  {Wise} M.~W.,    {David} L.~P.,  2005,
  \apj, 628, 629

\bibitem[\protect\citeauthoryear{{Nusser}, {Silk} \& {Babul}}{{Nusser}
  et~al.}{2006}]{2006MNRAS.373..739N}
{Nusser} A.,  {Silk} J.,    {Babul} A.,  2006, \mnras, 373, 739

\bibitem[\protect\citeauthoryear{{Osmond} \& {Ponman}}{{Osmond} \&
  {Ponman}}{2004}]{OP2004}
{Osmond} J.~P.~F.,  {Ponman} T.~J.,  2004, \mnras, 350, 1511

\bibitem[\protect\citeauthoryear{{Owen}, {Eilek} \& {Kassim}}{{Owen}
  et~al.}{2000}]{2000ApJ...543..611O}
{Owen} F.~N.,  {Eilek} J.~A.,    {Kassim} N.~E.,  2000, \apj, 543, 611

\bibitem[\protect\citeauthoryear{{Peterson}, {Paerels}, {Kaastra}, {Arnaud},
  {Reiprich}, {Fabian}, {Mushotzky}, {Jernigan} \& {Sakelliou}}{{Peterson}
  et~al.}{2001}]{2001A&A...365L.104P}
{Peterson} J.~R.,  {Paerels} F.~B.~S.,  {Kaastra} J.~S.,  {Arnaud} M.,
  {Reiprich} T.~H.,  {Fabian} A.~C.,  {Mushotzky} R.~F.,  {Jernigan} J.~G.,
  {Sakelliou} I.,  2001, \aap, 365, L104

\bibitem[\protect\citeauthoryear{{Popesso}, {Biviano}, {B{\" o}hringer},
  {Romaniello} \& {Voges}}{{Popesso} et~al.}{2005}]{2005AA...433..431P}
{Popesso} P.,  {Biviano} A.,  {B{\" o}hringer} H.,  {Romaniello} M.,    {Voges}
  W.,  2005, \aap, 433, 431

\bibitem[\protect\citeauthoryear{{Reynolds}, {Heinz} \& {Begelman}}{{Reynolds}
  et~al.}{2002}]{2002MNRAS.332..271R}
{Reynolds} C.~S.,  {Heinz} S.,    {Begelman} M.~C.,  2002, \mnras, 332, 271

\bibitem[\protect\citeauthoryear{{Sakelliou}, {Peterson}, {Tamura}, {Paerels},
  {Kaastra}, {Belsole}, {B{\"o}hringer}, {Branduardi-Raymont}, {Ferrigno}, {den
  Herder}, {Kennea}, {Mushotzky}, {Vestrand} \& {Worrall}}{{Sakelliou}
  et~al.}{2002}]{2002A&A...391..903S}
{Sakelliou} I.,  {Peterson} J.~R.,  {Tamura} T.,  {Paerels} F.~B.~S.,
  {Kaastra} J.~S.,  {Belsole} E.,  {B{\"o}hringer} H.,  {Branduardi-Raymont}
  G.,  {Ferrigno} C.,  {den Herder} J.~W.,  {Kennea} J.,  {Mushotzky} R.~F.,
  {Vestrand} W.~T.,    {Worrall} D.~M.,  2002, \aap, 391, 903

\bibitem[\protect\citeauthoryear{{Sandage}}{{Sandage}}{1978}]{1978AJ.....83..9%
04S}
{Sandage} A.,  1978, \aj, 83, 904

\bibitem[\protect\citeauthoryear{{Sanderson}, {Ponman}, {Finoguenov},
  {Lloyd-Davies} \& {Markevitch}}{{Sanderson}
  et~al.}{2003}]{2003MNRAS.340..989S}
{Sanderson} A.~J.~R.,  {Ponman} T.~J.,  {Finoguenov} A.,  {Lloyd-Davies} E.~J.,
     {Markevitch} M.,  2003, \mnras, 340, 989

\bibitem[\protect\citeauthoryear{{Sarazin}, {Rood} \& {Struble}}{{Sarazin}
  et~al.}{1982}]{1982A&A...108L...7S}
{Sarazin} C.~L.,  {Rood} H.~J.,    {Struble} M.~F.,  1982, \aap, 108, L7

\bibitem[\protect\citeauthoryear{{Struble} \& {Rood}}{{Struble} \&
  {Rood}}{1987}]{1987ApJS...63..543S}
{Struble} M.~F.,  {Rood} H.~J.,  1987, \apjs, 63, 543

\bibitem[\protect\citeauthoryear{{Struble} \& {Rood}}{{Struble} \&
  {Rood}}{1999}]{1999ApJS..125...35S}
{Struble} M.~F.,  {Rood} H.~J.,  1999, \apjs, 125, 35

\bibitem[\protect\citeauthoryear{{Willis}, {Pacaud}, {Valtchanov}, {Pierre},
  {Ponman}, {Read}, {Andreon}, {Altieri}, {Quintana}, {Dos Santos},
  {Birkinshaw}, {Bremer}, {Duc}, {Galaz}, {Gosset}, {Jones} \&
  {Surdej}}{{Willis} et~al.}{2005}]{2005MNRAS.363..675W}
{Willis} J.~P.,  {Pacaud} F.,  {Valtchanov} I.,  {Pierre} M.,  {Ponman} T.,
  {Read} A.,  {Andreon} S.,  {Altieri} B.,  {Quintana} H.,  {Dos Santos} S.,
  {Birkinshaw} M.,  {Bremer} M.,  {Duc} P.-A.,  {Galaz} G.,  {Gosset} E.,
  {Jones} L.,    {Surdej} J.,  2005, \mnras, 363, 675

\bibitem[\protect\citeauthoryear{{Wise}, {McNamara}, {Nulsen}, {Houck} \&
  {David}}{{Wise} et~al.}{2007}]{2007ApJ...659.1153W}
{Wise} M.~W.,  {McNamara} B.~R.,  {Nulsen} P.~E.~J.,  {Houck} J.~C.,    {David}
  L.~P.,  2007, \apj, 659, 1153

\end{thebibliography}
\label{lastpage} \end{document}